# Electrical manipulation of plasmon-phonon polaritons in heterostructures of graphene on biaxial crystals

*Zhiyong Li,*[1,2,3,4*] *Zefeng Chen,*[2,5] *Jingwen Ma,*[2] *Xin Hu,*[5] *Pan Wang,*[1,3,4] *Yungui Ma,*[1] *Jian-Bin Xu*[2*]

[1] State Key Laboratory of Modern Optical Instrumentation, College of Optical Science and Engineering, Zhejiang University, Hangzhou 310027, China.

[2] Department of Electronic Engineering, The Chinese University of Hong Kong, Hong Kong SAR 999077, China.

[3] Jiaxing Key Laboratory of Photonic Sensing & Intelligent Imaging, Jiaxing 314000, China.

[4] Intelligent Optics & Photonics Research Center, Jiaxing Research Institute Zhejiang University, Jiaxing 314000, China.

[5] School of Optoelectronic Science and Engineering, Soochow University, Suzhou 215006, China.

*Corresponding author. Email: lizhiyong@zju.edu.cn (Z.Y.L.); jbxu@ee.cuhk.edu.hk (J.-B.X.)

**Abstract**

Phonon polaritons in natural anisotropic crystals hold great promise for infrared nano-optics. However, the direct electrical control of these polaritons is difficult, preventing the development of active polaritonic devices. Here we propose the heterostructures of graphene on a biaxial crystal (α-phase molybdenum trioxide) slab and theoretically study the hybridized plasmon-phonon polaritons with dependence on the Fermi level of graphene from three aspects: dispersion relationships, iso-frequency contours, and the quantum spin Hall effects. We demonstrate the distinct wavelength tunability of the plasmon-phonon polaritons modes and the optical topologic transitions from open (hyperbolic) to closed (bow-tie-like) iso-frequency contours as the increase of the Fermi level of graphene. Furthermore, we observe the tunable quantum spin Hall effects of the plasmon-phonon polaritons, manifesting propagation direction switching by the Fermi level tuning of the graphene. Our findings open opportunities for novel electrically tunable polaritonic devices and programmable quantum optical networks.

**Introduction**

Polaritons in van der Waals (vdW) nanomaterials have shown great ability to confine light at the deep subwavelength scale (*1, 2*), especially in the mid-infrared region where conventional plasmonic (gold, silver et.al.) waveguides can hardly achieve (*3, 4*). Typical examples are plasmon polaritons in graphene, phonon polaritons in hexagonal boron nitride (h-BN) (*5-9*), α-$MoO_3$ (*10-16*), and α-$V_2O_5$ (*17*). To develop active polaritonic devices, the manipulation of the polaritons is necessary. The intrinsic plasmon in graphene can be tuned by gating (*18*), providing a promising route for active plasmon polariton devices (*19, 20*). The phonon polaritons in hyperbolic slabs can be tuned by thickness (*5, 10*), polar substrates (*15, 16*), and intercalation of Na atoms (*17*). More interestingly, the optical topological transition, switching the iso-frequency contours (IFCs) of the phonon polariton from open hyperbolic to closed elliptic, can be realized by the mode coupling in α-$MoO_3$ layers by twisting the stacking angles of two α-$MoO_3$ crystal slabs (*12-14*). However, these control methods for the phonon polaritons lack reconfigurability and it is still challenging to realize dynamic control (*21, 22*), especially electrical tunability.

Another remarkable opportunity for manipulation of phonon polaritons is via the spin-orbit interactions (SOI) of light (*23*), which results in the transverse spin-momentum locking, namely the quantum spin Hall effect of light (*24*). Circular polarization dipoles in the evanescence field are unidirectionally launched to the waveguide modes (*25, 26*) or metasurfaces (*27, 28*), showing great potential for chiral quantum optics (*29*) and integrated nano-optics. In the waveguides with ellipsoid 3D IFCs, the component directions of momentum and energy flows are the same (*25-28, 30*). However, they can be the opposite if the waveguides show hyperboloid 3D IFCs, and reversed unidirectional excitation of hyperbolic guided modes can be achieved (*31*). So, it is possible to switch the propagation directions of phonon polaritons by engineering the IFCs of surface optical modes.

In this work, we propose the graphene/α-$MoO_3$ heterostructures and theoretically demonstrate the electrical manipulation of the hybridized plasmon-phonon polaritons. Analytical and numerical calculation results show that the wavelength of hybridized surface plasmon-phonon polaritons ($SP^3$) and hyperbolic plasmon-phonon polaritons ($HP^3$) modes in the heterostructures can be distinctly tuned by graphene's Fermi level. Moreover, the topological transitions occur from open (hyperbolic) to closed (bow-tie-like) IFCs as the increase of graphene's Fermi level, which is attributed to the fusion of hyperbolic $HP^3$ and elliptical $SP^3$ modes at the interface. Furthermore, the topological transition of IFCs provides a way to tune the quantum spin Hall effects of the plasmon-phonon polaritons. Simulation results show that the energy flow direction at the interface can be switched by tuning the Fermi level of graphene, which is attributed to the competitive and opposite unidirectional excitation of $HP^3$ and $SP^3$modes by circular polarization dipoles. We believe our findings open great opportunities for electrical control of plasmon-phonon polaritons and will facilitate the development of novel programmable polaritonic devices.

## Results

### Plasmon-phonon polaritons in graphene/α-$MoO_3$ heterostructure

The optical response of graphene can be described with a surface conductivity model (*32*) considering the intraband and interband transition, with relaxation time $\tau$ =0.5 ps. The α-$MoO_3$ crystal is biaxial and poses three Reststrahlen bands in the mid-infrared range, corresponding to the three crystalline axes (*10, 11, 33*). The band 1, 2, and 3 of ranges from 545 to 851 $cm^{-1}$, from 820 to 972 $cm^{-1}$, and from 958 to 1010 $cm^{-1}$ are along the [001] ($y$-axis), [100] ($x$-axis), and [010] ($z$-axis) directions, respectively. The real parts of the corresponding permittivity components are negative, resulting in the hyperbolic phonon polariton ($HP^2$) responses. Lorentz model can be used for the approximation of the permittivity of α-$MoO_3$ (*11, 33*). More details of the optical response of graphene and the α-$MoO_3$ are given in sections S1 and S2.

To elaborate on the hybridization of plasmon-phonon polaritons in the graphene/α-$MoO_3$ heterostructure, we first consider the frequency/momentum dispersion relations of the two constituent elements (graphene and α-$MoO_3$). Figure 1(A-C) illustrates the cross profile of the three configurations: freestanding graphene, α-$MoO_3$ on $SiO_2$ substrate, and graphene/α-$MoO_3$ heterostructure on $SiO_2$ substrate. The dispersion relations can be plotted by the pseudo-colored map of the imaginary parts of the $p$-polarization Fresnel reflection coefficient $r_p$. Figure 1D shows the dispersion relations of the surface plasmon polariton ($SP^2$) of the freestanding graphene with Fermi levels of 0.15(a), 0.25(b), and 0.35(c) eV, exhibiting distinct tunability on Fermi levels. The dashed lines are described by the equation $q = 2i\varepsilon_0\omega / \sigma$ , where $\sigma$ is the surface conductivity of graphene. For the 100 nm-thick α-$MoO_3$ slab on the $SiO_2$ substrate, Figure 1E shows the dispersion relations of the $HP^2$ modes. In most parts of the band 1 and 2, type II $HP^2$ response with $\varepsilon_x > 0, \varepsilon_y < 0, \varepsilon_z > 0$ or $\varepsilon_x < 0, \varepsilon_y > 0, \varepsilon_z > 0$ exhibits single-direction propagation along [001] (*y*-axis) or [100] (*x*-axis) directions, respectively. In band 3, the type I $HP^2$ response with $\varepsilon_x > 0, \varepsilon_y > 0, \varepsilon_z < 0$ exhibits propagation along both the [100] (*x*-axis) and [001] (*y*-axis) directions. The results are consistent with the previous works (*11, 33*).

The in-plane permittivity $\varepsilon_2^t$ of α-$MoO_3$ along directions with the azimuthal angle $\varphi$ relative to the [100] (*x*-axis) direction can be calculated by $\varepsilon_2^t(\varphi) = \varepsilon_2^x \cos^2\varphi + \varepsilon_2^y \sin^2\varphi$, where $\varepsilon_2^x(\omega)$ and $\varepsilon_2^y(\omega)$ are the relative permittivity of the α-$MoO_3$ crystal along the [100] (*x*-axis) and [001] (*y*-axis) directions, respectively. For the configuration of graphene/α-$MoO_3$ heterostructure on $SiO_2$ substrate with polariton momentum $q(\omega,\varphi) >> \omega / c$, we derived the analytical formula for the polariton dispersion (See section S3):

$$q(\omega,\varphi) = \begin{cases} \dfrac{1}{d}\sqrt{-\dfrac{\varepsilon_2^z}{\varepsilon_2^t}}\left[m\pi + \varphi_1 + \varphi_2\right] & \varepsilon_2^t < 0, \varepsilon_2^z > 0 \\ \dfrac{1}{d}\sqrt{-\dfrac{\varepsilon_2^z}{\varepsilon_2^t}}\left[m\pi - \varphi_1 - \varphi_2\right] & \varepsilon_2^t > 0, \varepsilon_2^z < 0 \\ \dfrac{\pi\varepsilon_0\hbar^2\omega^2\left(\varepsilon_1 + \sqrt{\varepsilon_2^t\varepsilon_2^z}\right)}{e^2 E_F} & \varepsilon_2^t > 0, \varepsilon_2^z > 0 \end{cases} \tag{1}$$

$$\varphi_1 = \arctan\left(\frac{\varepsilon_1 - \dfrac{e^2 E_F q(\omega,\varphi)}{\pi\varepsilon_0\hbar^2\omega^2}}{\sqrt{-\varepsilon_2^t\varepsilon_2^z}}\right), \quad \varphi_2 = \arctan\left(\frac{\varepsilon_3}{\sqrt{-\varepsilon_2^t\varepsilon_2^z}}\right) \tag{2}$$

where $\varepsilon_0$, $e$ and $\hbar$ are the vacuum permittivity, electron charge, and reduced Plank constant, respectively; $\omega$, $d$ and $E_F$ are the angular frequency, the thickness of the α-$MoO_3$ slab, and the Fermi level of graphene, respectively; $\varepsilon_1(\omega)$, $\varepsilon_2^z(\omega)$ and $\varepsilon_3(\omega)$ are the relative permittivities of air, α-$MoO_3$ along the [010] ($z$-axis) direction, and $SiO_2$ substrate, respectively. If $\varepsilon_2^t\varepsilon_2^z<0$, the $HP^3$ modes inheriting from the $HP^2$ modes in α-$MoO_3$ slabs are supported in the heterostructure. If $\varepsilon_2^t>0$, $\varepsilon_2^z>0$, the $HP^3$ modes are no longer supported, and $SP^3$ modes inheriting from the $SP^2$ modes of graphene arise. Figure 1F plots the dispersion of the graphene/α-$MoO_3$ on $SiO_2$ substrate along [001] ($y$-axis) or [100] ($x$-axis) directions by mapping the imaginary parts of $r_p$ (See section S3). The blue dashed lines are analytically calculated modes ($m$ = 0, 1, 2, 3, 4) of Eq. (1) with $E_F$ = 0.35 eV and $\varepsilon_2^t=\varepsilon_2^x$ or $\varepsilon_2^t=\varepsilon_2^y$ for the [100] ($x$-axis) or [001] ($y$-axis) directions, respectively. The analytical dispersion curves agree well with the mapping of the imaginary parts of $r_p$. Compared to the type I and type II $HP^2$ of the α-$MoO_3$ slab, the type I and type II $HP^3$ in the heterostructure exhibit an obvious shift of momenta (at a fixed frequency), which are negative for the type II and positive for the type I. This behavior is similar to the graphene/hBN heterostructure (*34*) and can be seen from the $HP^3$ modes in Eq.(1). More interestingly, in the forbidden directions of bands 1 and 2, there are new $SP^3$ branches due to the introduction of graphene, which exhibits closed elliptical IFCs and contributes to the topological transitions discussed later. The Fermi level-dependent wavelength of $HP^3$ and $SP^3$ modes at the frequencies of 623 and 925 $cm^{-1}$ are shown in Figure S3, manifesting tunability and linear relationships.

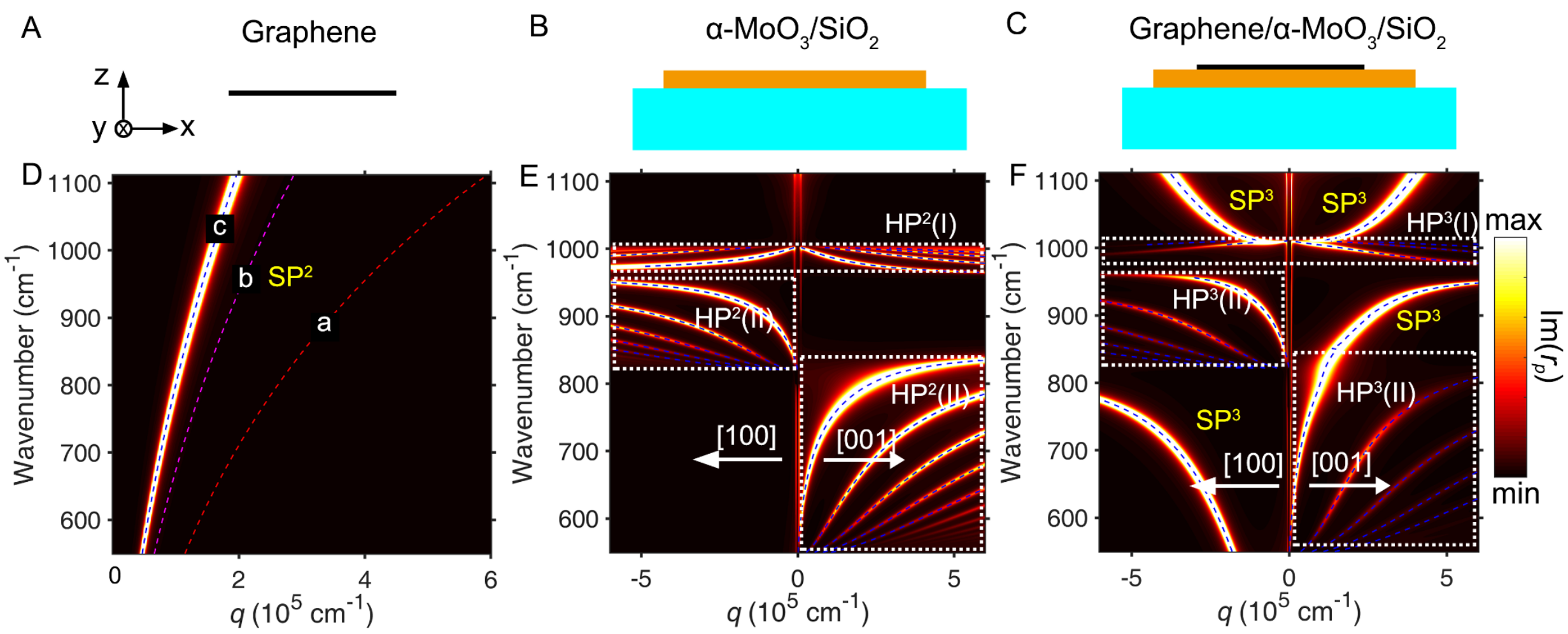

**Figure 1.** Dispersion relationships of the three structures of freestanding graphene, α-$MoO_3$/$SiO_2$, and graphene/α-$MoO_3$/$SiO_2$. (A-C) Schematic illustration of the three structures, the coordinate in (A) applies to the entire article. (D) Calculated dispersion of the surface plasmon polaritons ($SP^2$) in freestanding graphene with the Fermi level $E_F$ = 0.15 (a), 0.25 (b), and 0.35 (c) eV. (E) Calculated dispersion of the hyperbolic phonon polaritons ($HP^2$) along [100] (*x*-axis) and [001] (*y*-axis) directions supported by a 100 nm-thick α-$MoO_3$ slab on $SiO_2$ substrate. (F) Calculated dispersion of the surface plasmon-phonon polaritons ($SP^3$) and hyperbolic plasmon-phonon polaritons ($HP^3$) supported by a graphene/α-$MoO_3$ heterostructure on $SiO_2$ substrate. The thickness of the α-$MoO_3$ slab is 100 nm and the Fermi level of graphene is $E_F$ = 0.35 eV. The pseudo-colored maps in (D-F) describe the imaginary parts of the reflection coefficient $r_p$, and the dashed lines describe the analytically calculated dispersions of polaritons.

## Electrically tunable topological transitions of plasmon-phonon polaritons

Topological transitions of the polaritonic IFCs have been studied in the twisted bi-layer α-$MoO_3$ slabs (*13, 14, 35*) and plasmonic hyperbolic metasurfaces (*36, 37*), where the twisted angle is crucial for the transition. However, dynamic control of the topologic transitions is still challenging (*21, 22*). Here we predict electrically tunable topological transitions of plasmon-phonon polariton in the graphene/α-$MoO_3$ heterostructures by tuning the Fermi levels of graphene. Without losing any generality, we study three typical frequencies of 623, 925, and 1000 $cm^{-1}$, which are in the band 1, 2, and 3 of the permittivity of $\alpha$-$MoO_3$, respectively. We numerically simulate the propagation of polaritons with a $z$-oriented dipole excitation $\mathbf{p} = [0, 0, 1]$, located 60 nm above the graphene monolayer. Figure 2(A-D) and (I-L) plot the electric field distribution of the real parts of $z$-components, namely Re($E_z$), without graphene and with Fermi levels of 0.15, 0.25, and 0.35 eV of graphene, respectively. The frequencies of Figure 2 (A-D) and (I-L) are 623 and 925 $cm^{-1}$, respectively. With the introduction of graphene and the increase of Fermi level, the ray-like propagation of $HP^2$ and $HP^3$ are destructed by mixing with $SP^3$ propagating along their forbidden directions. We conduct Fourier transform on the electric field Re($E_z$), and obtain the corresponding in-plane IFCs as shown in Figures 2 (E-H) and (M-P). The dashed lines are analytically calculated from Eq. (1) by solving the momentum $q(\varphi)$ at the fixed frequencies and mode orders ($m$ = 0, 1, 2), agreeing well with the full-wave simulation results. Remarkably, with the increase of Fermi levels, the IFCs gradually evolve from open hyperbolic to closed bow-tie-like contours at frequencies of 623 and 925 $cm^{-1}$. In Figure S5, we plot the evolutions of IFCs by mapping the imaginary parts of $r_p(q,\varphi)$ at the frequencies of 625, 925, and 1000 $cm^{-1}$, respectively, again with good consistency with the analytical results from Eq. (1). And there is no topological transition at 1000 $cm^{-1}$ because both $SP^3$ and type I $HP^3$ exhibit closed elliptic in-plane IFCs at this frequency.

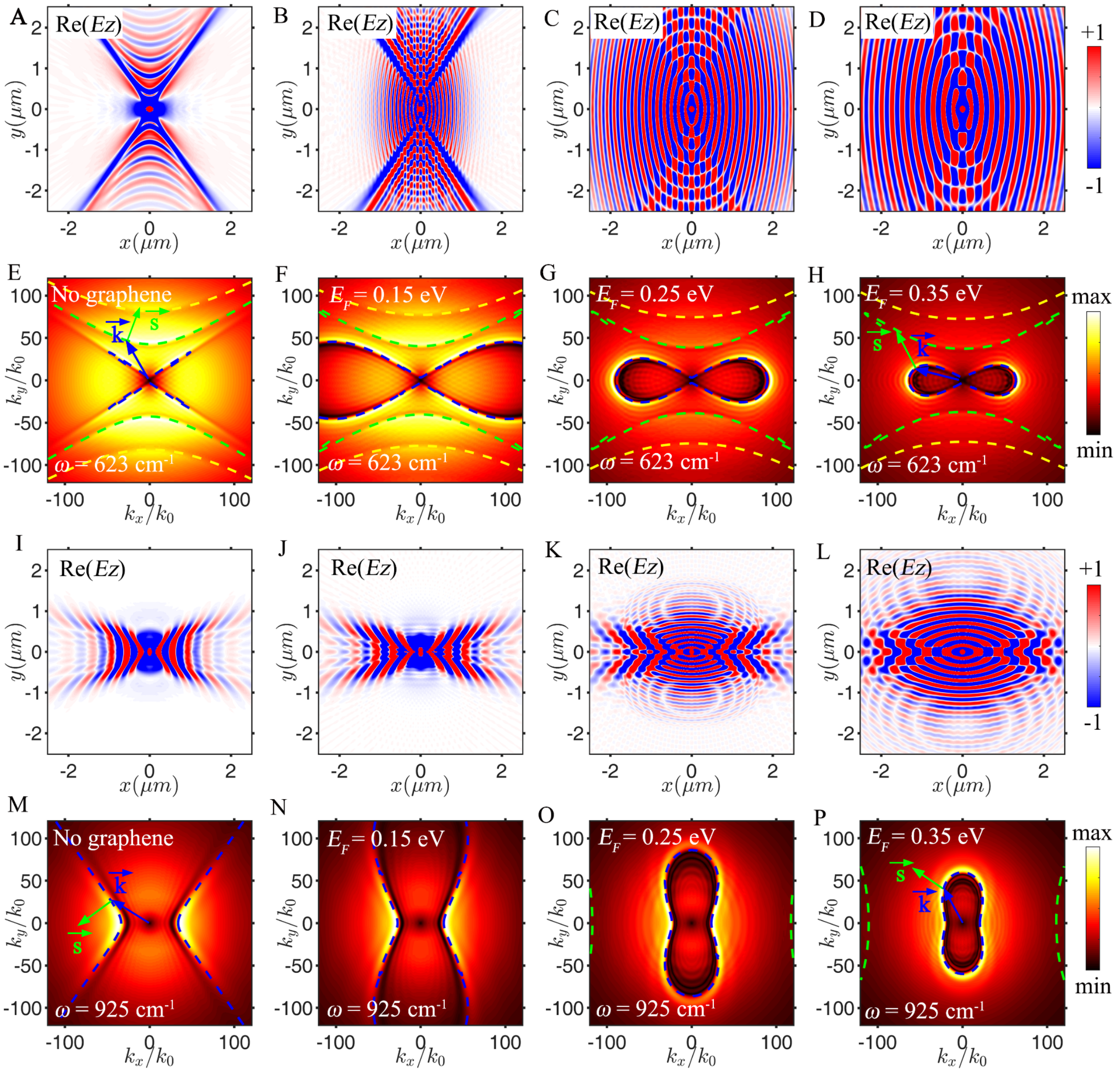

**Figure 2.** Electrically tunable topological transitions of plasmon-phonon polaritons in the graphene/α-$MoO_3$/$SiO_2$ structure. The dipole excitation **p** = [0, 0 ,1] was located 60-nm-above the graphene. (A-D, I-L) Calculated real parts of z-components of the electric field distribution. (E-H, M-P) Corresponding Fourier transform calculated IFCs without graphene and with graphene at the varying Fermi level. The blue dashed lines are the calculated $HP^3$ modes of $m$=0 and $SP^3$ modes, and the green and yellow dashed lines are the $HP^3$ modes of $m$ = 1 and $m$ = 2, respectively. The blue and green arrows in (E, H, M, P) illustrate the momentum (**k**) and Poynting vector (**S**) directions. The frequency for (A-H) is 623 $cm^{-1}$, while for (I-P) is 925 $cm^{-1}$. The thickness of the α-$MoO_3$ slab is 100 nm.

**Electrically tunable quantum spin Hall effect of plasmon-phonon polaritons**

Unidirectional routing of light is important for photonic devices and quantum information processing (*29, 38*). A well-known approach to achieve unidirectional propagation is to break the reciprocity with a magnetic field (*38, 39*), but it is impractical for compact devices owning to the limited magnetic response in most optical media. Recently, the photonic quantum spin Hall effect based on spin-orbit interaction(SOI) of light has been demonstrated for the unidirectional excitation of guiding modes (*25-27*) and valley excitons separation (*28, 30, 40*), providing us a new alternative.

Near-field excitation of the investigated waveguides with a circular electric dipole is used for the study of photonic SOI (*25-28, 30, 36*). It is convenient to analyze the unidirectional excitation by decomposing the field of the dipole into spatial-frequency components $k_x$ and $k_y$ (*25*). As shown in Figure S5(A-D) and (E-H), circular dipoles $\mathbf{p}$ = [1, 0, −$i$] and $\mathbf{p}$ = [0, 1, −$i$] exhibit highly asymmetric magnetic field distribution along $k_x$ and $k_y$ directions, respectively. The asymmetric distributions will lead to predominant excitation directions of the waveguides by the circular dipoles, depending on the 3D IFCs (ellipsoid or hyperboloid) and in-plane IFCs (See section S6). Specifically, Figure 2(E, H, M, P) illustrate the directions of a specific momentum $\mathbf{k}$ and the corresponding Poynting vectors $\mathbf{S}$ for the corresponding in-plane IFCs of the polariton modes. For the frequency of 623 $cm^{-1}$, the directions of the $x$ components of the Poynting vector $\mathbf{S}$ switch from positive at no graphene to negative at $E_F$ = 0.35 eV of graphene, which means the excitation direction switching along the $x$-axis. However, the corresponding $y$ components remain the same as positive, and no direction switching. Similar results can be obtained for the frequency of 925 $cm^{-1}$.

To verify our prediction and investigate the quantum spin Hall effect of the plasmon-phonon polaritons in the single α-$MoO_3$ slab and graphene/α-$MoO_3$ heterostructure, we numerically simulated the wave propagation of the structures with circular dipoles excitation. Figure 3A and 3E show the Re($E_z$) and Poynting vector magnitude (abs(**S**)) distribution of a 100 nm α-$MoO_3$ slab at $SiO_2$ substrate at the frequency of 623 $cm^{-1}$, excited by a circular dipole $\mathbf{p} = [1, 0, -i]$. The dipole $\mathbf{p} = [1, 0, -i]$ poses a much higher magnetic field in the $k_x$ direction of negative than positive, resulting in the predominant excitation of the $HP^2$ mode towards the positive $x$-axis direction, which is attributed to the reversed directions of the $x$-components of momentum and Poynting vector as shown in Figure 2E. However, the excitation of the $HP^2$ mode at the frequency of 925 $cm^{-1}$ exhibits priority towards the negative $x$-axis direction as shown in Figures 3I and 3M, which is attributed to the same directions of the $x$-components of momentum and Poynting vector as shown in Figure 2M. Furthermore, we numerically simulated the wave propagation of the graphene/α-$MoO_3$ heterostructure at the $SiO_2$ substrate with different Fermi levels of graphene. Figure 3(B-D) and 3(F-H) plot the Re($E_z$) and Poynting vector magnitude distribution at the frequency of 623 $cm^{-1}$ with Fermi levels of 0.15, 0.25, and 0.35 eV. As the increase of the Fermi levels, the $SP^3$ modes gradually dominate the surface wave propagation, resulting in the energy flow switching from the predominant positive of the $HP^2$ or $HP^3$ modes to the negative of the $SP^3$ modes. For the frequency of 925 $cm^{-1}$ as shown in Figure 3(J-L, N-P), although the $SP^3$ gradually dominates the propagation as the increase of Fermi levels, predominant negative energy flow remains, which is consistent with the illustrated negative direction of $x$-components of Poynting vectors in Figure 2M and 2P. More simulation results of the surface wave propagation at the frequency of 623 and 925 $cm^{-1}$ by excitation of dipole $\mathbf{p} = [0, 1, -i]$ are shown in Figure S6. Figure

S7 shows the wave propagation at 1000 cm$^{-1}$ by excitation of the dipole $\mathbf{p} = [1, 0, -i]$ or $\mathbf{p} = [0, 1, -i]$. Energy flow switches of the unidirectional excitation can also be observed.

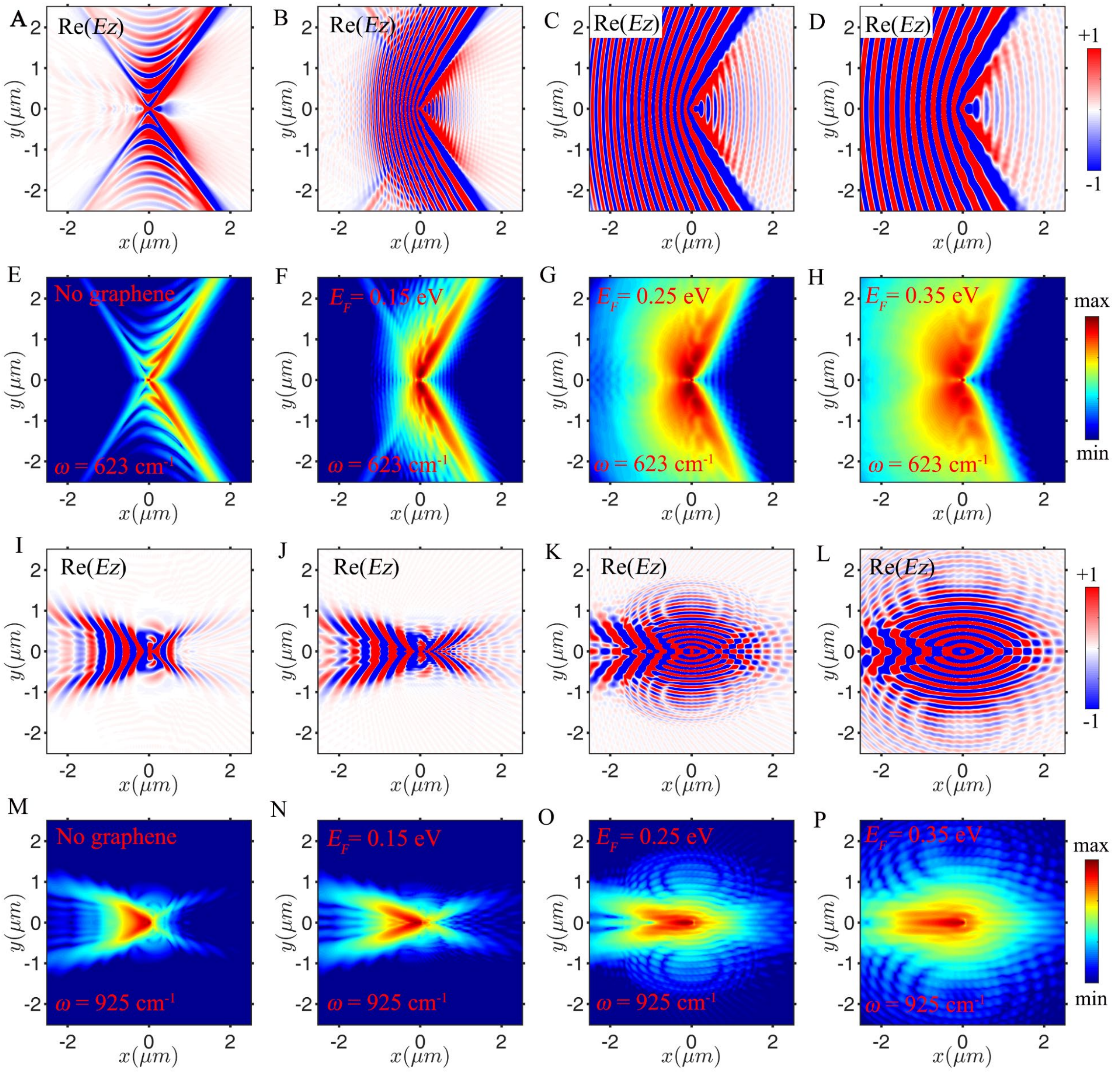

**Figure 3**. Electrically tunable spin Hall effect of plasmon-phonon polaritons in the graphene/α-$MoO_3$/$SiO_2$ structure. The dipole excitation $\mathbf{p} = [1, 0, -i]$ was located 60-nm-above the graphene. (A-D, I-L) Calculated real parts of $z$-components of the electric field distribution. (E-H, M-P) Log scale plots of the corresponding Poynting vector magnitudes (abs(**S**)) without graphene and with graphene at the varying Fermi level. The frequency for (A-H) is 623 cm$^{-1}$, while for (I-P) is 925 cm$^{-1}$. The thickness of the α-$MoO_3$ slab is 100 nm.

To quantitively describe the electrically tunable quantum spin Hall effect of plasmon-phonon polaritons in the graphene/α-$MoO_3$ heterostructure, we calculate the directional factors defined below:

$$D_0 = \frac{I_+ - I_-}{I_+ + I_-} \tag{3}$$

where $I_+$ and $I_-$ are the integrals of the Poynting vector magnitude at the positive half-space ($x >$ 0 or $y > 0$) and the negative half-space ($x < 0$ or $y < 0$), respectively. Figure 4 plot the directional factors as a function of the Fermi levels at frequency of 623 and 925 $cm^{-1}$, by excitation of circular dipole $\mathbf{p} = [1, 0, -i]$ or $\mathbf{p} = [0, 1, -i]$. As the variation of the Fermi levels, the directional factors exhibit remarkable tunability from positive to negative with a proper circular dipole excitation, indicating a great potential for the electrically tunable polaritonic devices. The electrical tunability may also provide new freedom beyond spin for the unidirectional routings.

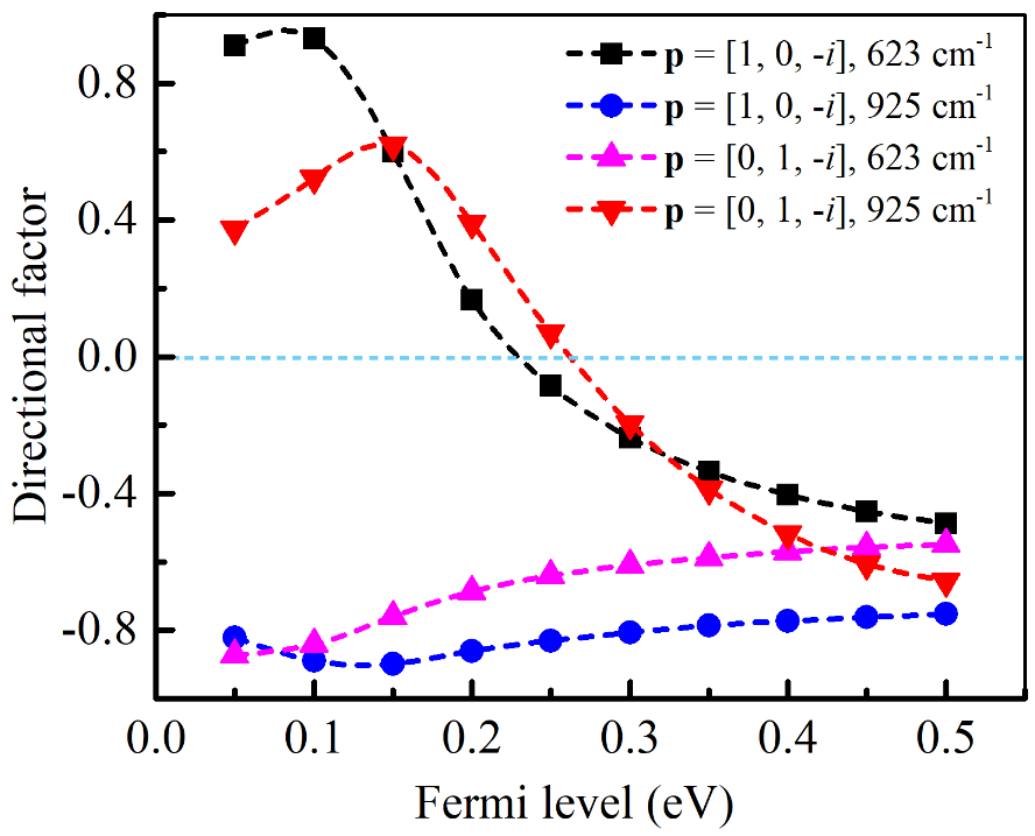

**Figure 4.** Directional factors of the plasmon-phonon polaritons in the graphene/α-$MoO_3$/$SiO_2$ structure with excitations of circular dipoles. The solid symbols are calculated values from the simulation results, and the black, blue, magenta and red dashed lines are the fittings showing the variation tendency. The horizontal cyan dashed line corresponds to $D_0 = 0$.

## Discussion

In this work, we have investigated the hybridization of plasmon and phonon polaritons in the graphene/α-$MoO_3$ heterostructure and revealed the electrically tunable topological transitions of the in-plane IFCs from open hyperbolic to closed bow-tie-like curve. We have also numerically studied the electrical tunability of the polaritonic spin Hall effects in the heterostructure, indicating new freedom for unidirectional routings and a great potential for novel polaritonic devices. Further optimization of the unidirectional routing can be achieved by using a proper elliptical dipole excitation (*25, 36, 41*). Also, excitation with Janus dipole or Huygens dipole may provide new ways for directionality control beyond the spin-momentum locking mechanism (*42*). Besides, the study on the local density of states (*43*), and strong coupling between plasmon and phonon polaritons (*44*) in the graphene/α-$MoO_3$ heterostructure are also desirable. At last, we expect the experimental demonstration of our findings in the near field or by far-field methods.

## Materials and Methods

### Full-wave numerical simulations

We performed the full-wave simulations using the commercially available software Lumerical FDTD solutions. The Lorentz model with fitted parameters was used for the permittivity of α-$MoO_3$ as described in section S2. The built-in surface conductivity model of graphene with the specified Fermi level was used to be on the top of the α-$MoO_3$, with a relaxation time of 0.5 ps. The permittivity of the $SiO_2$ substrate comes from Palik's data (*45*). The excitation dipole of $\mathbf{p}$ = [0, 0, 1], or $\mathbf{p}$ = [1, 0, −$i$], or $\mathbf{p}$ = [0, 1, −$i$] was located 60-nm-above the graphene. The perfectly matched layer (PML) boundary conditions were used along the x, y, and z-axis. The electric field monitor is placed on the top of the α-MoO3 slab, and the IFCs were calculated by the Fourier transform of the real parts of the Ez of the monitors.

**Acknowledgments**

The work is in part supported by the Research Grants Council of Hong Kong, particularly, via Grant Nos. AoE/P-701/20, 14203018, and CUHK Group Research Scheme, National Natural Science Foundation of China 62075196, and Natural Science Foundation of Zhejiang Province LXZ22F050001.

**Author Contributions**

Z.Y.L. and J.-B.X. conceived the concept. Z.Y.L. performed the simulations, calculations, and analysis. Z.F.C. and J.W.M. assisted in the simulations and analysis. J.-B.X supervised the work. Z.Y.L. wrote the original draft of the manuscript. All authors reviewed and contributed to the manuscripts.

**Competing interests**

The authors declare no competing interests.

# Supporting Information of

# "Electrical manipulation of plasmon-phonon polaritons in heterostructures of graphene on biaxial crystals"

*Zhiyong Li,*[1,2,3,4*] *Zefeng Chen,*[2,5] *Jingwen Ma,*[2] *Xin Hu,*[5] *Pan Wang,*[1,3,4] *Yungui Ma,*[1] *Jian-Bin Xu*[2*]

[1] State Key Laboratory of Modern Optical Instrumentation, College of Optical Science and Engineering, Zhejiang University, Hangzhou 310027, China.

[2] Department of Electronic Engineering, The Chinese University of Hong Kong, Hong Kong SAR 999077, China.

[3] Jiaxing Key Laboratory of Photonic Sensing & Intelligent Imaging, Jiaxing 314000, China.

[4] Intelligent Optics & Photonics Research Center, Jiaxing Research Institute Zhejiang University, Jiaxing 314000, China.

[5] School of Optoelectronic Science and Engineering, Soochow University, Suzhou 215006, China.

*Corresponding author. Email: lizhiyong@zju.edu.cn (Z.Y.L.); jbxu@ee.cuhk.edu.hk (J.-B.X.)

**S1. Optical response of graphene**

In the FDTD simulation, we use the built-in surface conductivity model of graphene from the Kubo formula (*1*) including the intraband and interband transitions as below:

$$\sigma(\omega, E_F, \tau, T) = \sigma_{intra}(\omega, E_F, \tau, T) + \sigma_{inter}(\omega, E_F, \tau, T) \tag{S1}$$

$$\sigma_{intra}(\omega, E_F, \tau, T) = i\frac{e^2 k_B T}{\pi\hbar^2\left(\omega + i\tau^{-1}\right)}\left(\frac{E_F}{k_B T} + 2\ln\left(e^{-E_F/k_B T} + 1\right)\right) \tag{S2}$$

$$\sigma_{\text{inter}}(\omega, E_F, \tau, T) = i\frac{e^2(\omega + i\tau^{-1})}{\pi\hbar^2}\int_0^\infty \frac{f_d(-\xi) - f_d(\xi)}{(\omega + i\tau^{-1})^2 - 4(\xi/\hbar)^2} d\xi \tag{S3}$$

$$f_d(\xi) = \frac{1}{\exp((\xi - E_F)/k_B T) + 1} \tag{S4}$$

Here $\omega$ is the angular frequency, $E_F$ is the Fermi energy, $\tau$ is the relaxation time, $T$ is the temperature, $e$ is the electron charge, $\hbar$ is the reduced Plank constant, $k_B$ is the Boltzmann constant. In the mid-infrared range, the intra-band transition dominates the response of graphene, then the surface conductivity can be simplified as follow when $E_F >> k_B T$ and $\omega >> \tau^{-1}$:

$$\sigma \simeq \frac{e^2 E_F}{\pi\hbar^2}\frac{i}{\omega + i\tau^{-1}} \simeq \frac{ie^2 E_F}{\pi\hbar^2\omega} \tag{S5}$$

### S2. The relative permittivity of α-$MoO_3$

By fitting the experimental results from reference (*2*) with the Lorentz model of equation S1 from the Lumerical FDTD solutions, we obtain the corresponding parameters for the *x* [100], *y* [001], and *z* [010] directions as shown in the Table S1. Figure S1 plots the real parts of the permittivity components been used for calculation, and the three Reststrahlen bands are pointed out.

$$\varepsilon_{total}(f) = \varepsilon + \frac{\varepsilon_{lorentz} \cdot \omega_0^2}{\omega_0^2 - 2i\delta_0 \cdot (2\pi f) - (2\pi f)^2} \tag{S6}$$

where $f$ is the frequency in the unit of *Hz*.

**Table S1.** Parameters used in the calculations of the permittivity of α-$MoO_3$

| | *x* [100] | *y* [001] | *z* [010] |
|---|---|---|---|
| permittivity $\varepsilon$ | 5.189 | 6.069 | 4.409 |
| Lorentz permittivity $\varepsilon_{lorentz}$ | 2.006 | 8.734 | 0.5161 |
| Lorentz resonance $\omega_0$ (rad/s) | $1.55\times10^{14}$ | $1.03\times10^{14}$ | $1.80\times10^{14}$ |
| Lorentz linewidth $\delta_0$ (rad/s) | $5.60\times10^{11}$ | $9.00\times10^{11}$ | $1.80\times10^{11}$ |

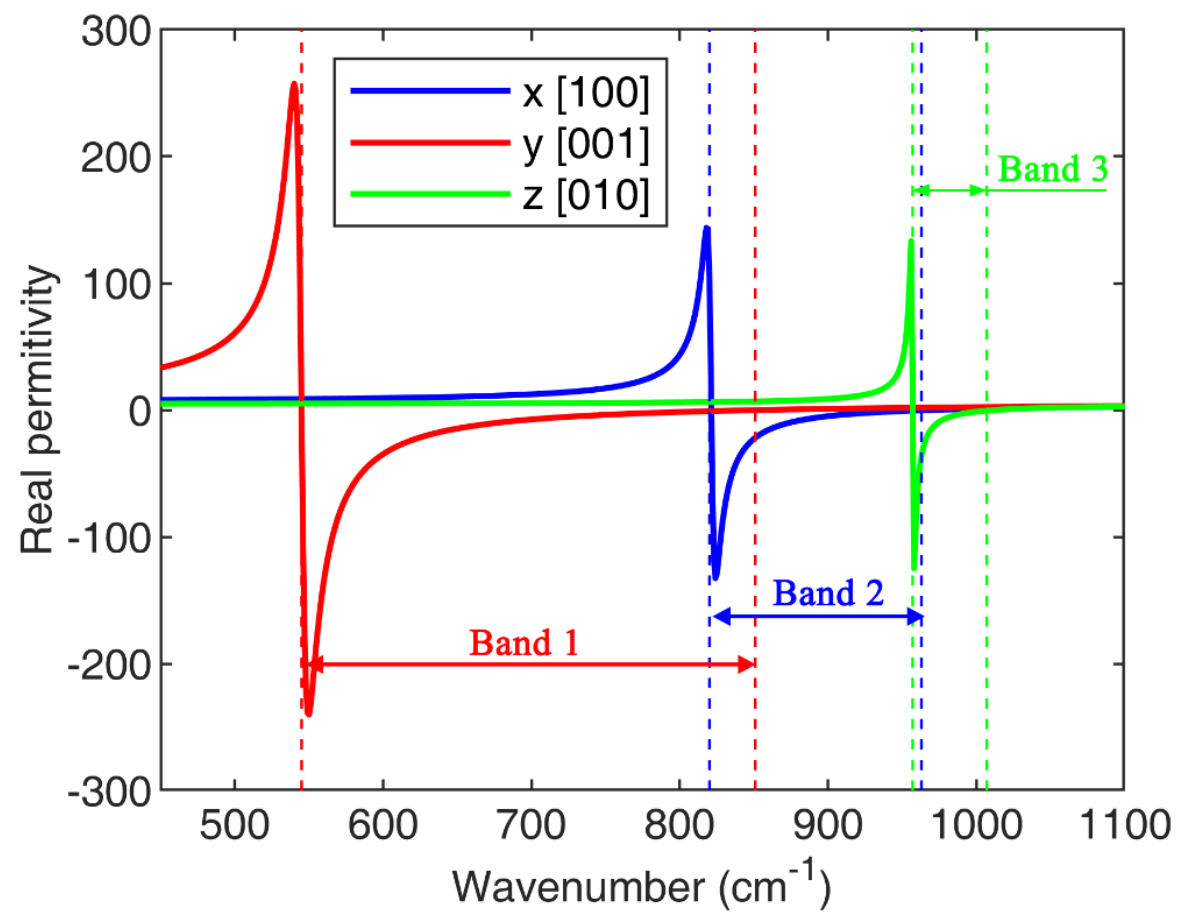

**Figure S1.** The real parts of the relative permittivity of α-$MoO_3$, the three Reststrahlen bands are pointed out.

## S3. Analytical model of the plasmon-phonon polaritons in graphene/α-$MoO_3$/$SiO_2$ heterostructures

The graphene/α-$MoO_3$/$SiO_2$ heterostructure is treated as an infinite stratified structure as illustrated in Figure S2. The graphene is located on the top of the α-$MoO_3$ slab. The permittivity of α-$MoO_3$ and surface conductivity of graphene are described in Section 1 and 2, respectively. The permittivity of $SiO_2$ is from Palik's data (*3*). The three regions of the air, α-$MoO_3$, and $SiO_2$ are represented by $j$ = 1, 2, and 3, respectively. The thickness of the α-$MoO_3$ slab is $d$.

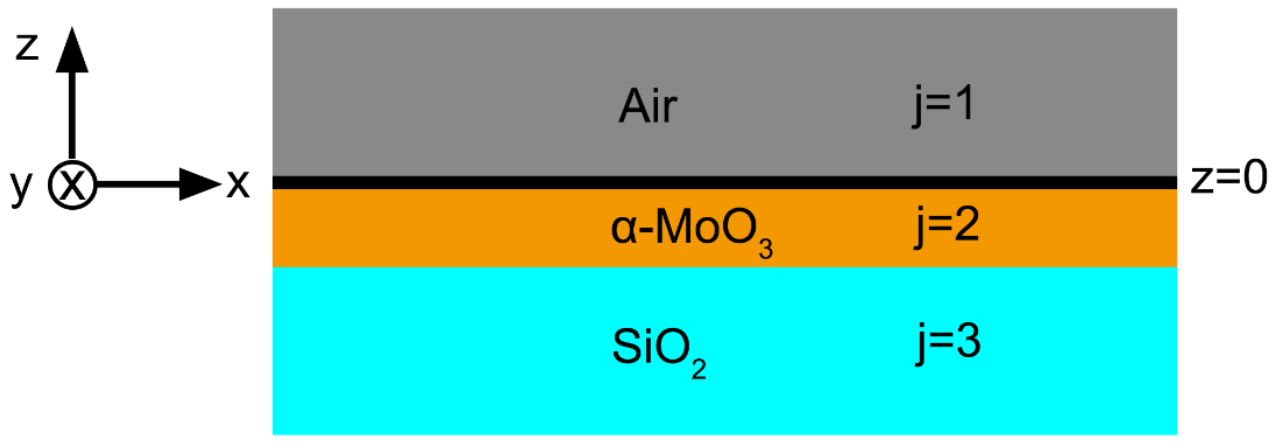

**Figure S2.** Schematic diagram of the graphene/α-$MoO_3$/$SiO_2$ heterostructure, the graphene is located on the top surface of the α-$MoO_3$ slab.

The effective in-plane permittivity of α-$MoO_3$ along arbitrary directions can be obtained by:

$$\varepsilon_2^t = \varepsilon_2^x \cos^2\varphi + \varepsilon_2^y \sin^2\varphi \tag{S7}$$

where $\varphi$ is the azimuthal angle relative to the $x$-axis.

For the $p$-polarization, the Fresnel reflection coefficient can be obtained by:

$$r_p = \frac{r_{12} + r_{23}(1 - r_{12} - r_{21})e^{2ik_2^z d}}{1 - r_{21} r_{23} e^{2ik_2^z d}} \tag{S8}$$

where

$$\begin{aligned} &r_{12} = \frac{Q_2 - Q_1 + S}{Q_2 + Q_1 + S}, \quad r_{21} = \frac{Q_1 - Q_2 + S}{Q_1 + Q_2 + S}, \quad r_{23} = \frac{Q_3 - Q_2}{Q_3 + Q_2}, \\ &\mathrm{Q}_j = \frac{\varepsilon_j^t}{k_j^z}, \quad \mathrm{S} = \frac{\sigma}{\varepsilon_0 \omega}, \quad k_j^z = \sqrt{\varepsilon_j^t \left(\frac{\omega^2}{c^2} - \frac{q^2}{\varepsilon_j^z}\right)} \end{aligned} \tag{S9}$$

and $\varepsilon_0$ is the vacuum permittivity, $\sigma$ is the surface conductivity of graphene, $\omega$ is the angular frequency, and $q$ is the in-plane momentum. By pseudo-colored mapping the imaginary parts of the Fresnel reflection coefficient $r_p(q,\omega)$ of Eq. (S8) with $\varepsilon_2^t = \varepsilon_2^x$ or $\varepsilon_2^t = \varepsilon_2^y$, we can obtain the dispersion relationships along the $x$-axis ([100]) or $y$-axis ([001]) directions, as shown in Figure 1F of the main text. For the case without graphene, we set $S$=0, then we obtain the dispersion relationships of α-$MoO_3$ slab on $SiO_2$ substrate as shown in Figure 1E of the main text. For the case of freestanding graphene, we calculate the Fresnel reflection coefficient $r_p = r_{12} = \frac{Q_2 - Q_1 + S}{Q_2 + Q_1 + S}$, and $\varepsilon_1 = \varepsilon_2 = 1$, then we obtain the dispersion relationships of freestanding graphene as shown in Figure 1D of the main text. Moreover, we can plot the in-plane IFCs at specific frequencies by calculating the Fresnel reflection coefficient $r_p(q,\varphi)$, as shown in Figure S4.

To further deeply understand the physics, we make the following analysis with two cases of HP[3] modes when $\varepsilon_2^t \varepsilon_2^z < 0$ and SP[3] modes when $\varepsilon_2^t > 0,\ \varepsilon_2^z > 0$. Here we ignore the imaginary parts of the permittivity of α-$MoO_3$ to simplify the derivations, but the physics remains.

1) HP[3] modes when $\varepsilon_2^t \varepsilon_2^z < 0$

Since we study $q \gg \frac{\omega}{c}$, From Eq.(S9) we can obtain:

$$
\begin{aligned}
& k_1^z \simeq iq,\ \ Q_1 \simeq -\frac{i\varepsilon_1}{q} \\
& k_2^z \simeq q\sqrt{-\frac{\varepsilon_2^t}{\varepsilon_2^z}},\ \ \mathrm{Q}_2 = \begin{cases} \frac{1}{q}\sqrt{-\varepsilon_2^t \varepsilon_2^z} & \varepsilon_2^t > 0, \varepsilon_2^z < 0 \\ -\frac{1}{q}\sqrt{-\varepsilon_2^t \varepsilon_2^z} & \varepsilon_2^t < 0, \varepsilon_2^z > 0 \end{cases} \\
& k_3^z \simeq iq,\ \ Q_3 \simeq -\frac{i\varepsilon_3}{q}
\end{aligned} \tag{S10}
$$

And we have $\varepsilon_2^t \varepsilon_2^z < 0$, so $k_2^z$ should be real, meaning propagating waveguide modes are possible if the standing wave conditions below are satisfied:

$$
2k_2^z d + \phi_{21} + \phi_{23} = 2m\pi,\ \ \text{m=0, 1, 2,}\cdots \tag{S11}
$$

where $d$ is the thickness of the α-$MoO_3$ slab, $\phi_{21}$ and $\phi_{23}$ are the phase changes of reflections at the interfaces of α-$MoO_3$/air and α-$MoO_3$/$SiO_2$, respectively.

By substituting Eq. (S10) into Eq. (S9), we can obtain:

$$
r_{21} = \begin{cases} \dfrac{-i\varepsilon_1 + \sqrt{-\varepsilon_2^t \varepsilon_2^z} + \dfrac{\sigma q}{\varepsilon_0 w}}{-i\varepsilon_1 - \sqrt{-\varepsilon_2^t \varepsilon_2^z} + \dfrac{\sigma q}{\varepsilon_0 w}} & \varepsilon_2^t < 0, \varepsilon_2^z > 0 \\ \dfrac{-i\varepsilon_1 - \sqrt{-\varepsilon_2^t \varepsilon_2^z} + \dfrac{\sigma q}{\varepsilon_0 w}}{-i\varepsilon_1 + \sqrt{-\varepsilon_2^t \varepsilon_2^z} + \dfrac{\sigma q}{\varepsilon_0 w}} & \varepsilon_2^t > 0, \varepsilon_2^z < 0 \end{cases} \tag{S12}
$$

$$r_{23}=\begin{cases}\dfrac{i\varepsilon_3-\sqrt{-\varepsilon_2^t\varepsilon_2^z}}{i\varepsilon_3+\sqrt{-\varepsilon_2^t\varepsilon_2^z}} & \varepsilon_2^t<0,\varepsilon_2^z>0\\ \dfrac{i\varepsilon_3+\sqrt{-\varepsilon_2^t\varepsilon_2^z}}{i\varepsilon_3-\sqrt{-\varepsilon_2^t\varepsilon_2^z}} & \varepsilon_2^t>0,\varepsilon_2^z<0\end{cases}\tag{S13}$$

In the mid-infrared region of our interest, we have $\Im(\sigma)\gg\Re(\sigma)$, so we ignore the real parts of the surface conductivity of the graphene. By using $\phi_{21}=\arg(\mathrm{r}_{21})$, $\phi_{23}=\arg(\mathrm{r}_{23})$, and then we obtain:

$$\phi_{21}=\begin{cases}2\arctan\left(\dfrac{\frac{\Im(\sigma)q}{\varepsilon_0\omega}-\varepsilon_1}{\sqrt{-\varepsilon_2^t\varepsilon_2^z}}\right)-\pi & \varepsilon_2^t<0,\varepsilon_2^z>0,\ \dfrac{\Im(\sigma)q}{\varepsilon_0\omega}>\varepsilon_1\\ 2\arctan\left(\dfrac{\frac{\Im(\sigma)q}{\varepsilon_0\omega}-\varepsilon_1}{\sqrt{-\varepsilon_2^t\varepsilon_2^z}}\right)+\pi & \varepsilon_2^t<0,\varepsilon_2^z>0,\ \dfrac{\Im(\sigma)q}{\varepsilon_0\omega}<\varepsilon_1\\ -2\arctan\left(\dfrac{\frac{\Im(\sigma)q}{\varepsilon_0\omega}-\varepsilon_1}{\sqrt{-\varepsilon_2^t\varepsilon_2^z}}\right)+\pi & \varepsilon_2^t>0,\varepsilon_2^z<0,\ \dfrac{\Im(\sigma)q}{\varepsilon_0\omega}>\varepsilon_1\\ -2\arctan\left(\dfrac{\frac{\Im(\sigma)q}{\varepsilon_0\omega}-\varepsilon_1}{\sqrt{-\varepsilon_2^t\varepsilon_2^z}}\right)-\pi & \varepsilon_2^t>0,\varepsilon_2^z<0,\ \dfrac{\Im(\sigma)q}{\varepsilon_0\omega}<\varepsilon_1\end{cases}\tag{S14}$$

$$\phi_{23}=\begin{cases}-2\arctan\left(\dfrac{\varepsilon_3}{\sqrt{-\varepsilon_2^t\varepsilon_2^z}}\right)+\pi & \varepsilon_2^t<0,\varepsilon_2^z>0\\ 2\arctan\left(\dfrac{\varepsilon_3}{\sqrt{-\varepsilon_2^t\varepsilon_2^z}}\right)-\pi & \varepsilon_2^t>0,\varepsilon_2^z<0\end{cases}\tag{S15}$$

By substituting Eq. (S14) and (S15) into Eq. (S11), we can obtain:

$$q=\begin{cases}\frac{1}{d}\sqrt{-\frac{\varepsilon_2^z}{\varepsilon_2^t}}\left[m\pi+\varphi_1+\varphi_2\right] & \varepsilon_2^t<0,\varepsilon_2^z>0,\ \frac{\Im(\sigma)q}{\varepsilon_0\omega}>\varepsilon_1\\ \frac{1}{d}\sqrt{-\frac{\varepsilon_2^z}{\varepsilon_2^t}}\left[(m-1)\pi+\varphi_1+\varphi_2\right] & \varepsilon_2^t<0,\varepsilon_2^z>0,\ \frac{\Im(\sigma)q}{\varepsilon_0\omega}<\varepsilon_1\\ \frac{1}{d}\sqrt{-\frac{\varepsilon_2^z}{\varepsilon_2^t}}\left[m\pi-\varphi_1-\varphi_2\right] & \varepsilon_2^t>0,\varepsilon_2^z<0,\ \frac{\Im(\sigma)q}{\varepsilon_0\omega}>\varepsilon_1\\ \frac{1}{d}\sqrt{-\frac{\varepsilon_2^z}{\varepsilon_2^t}}\left[(m+1)\pi-\varphi_1-\varphi_2\right] & \varepsilon_2^t>0,\varepsilon_2^z<0,\ \frac{\Im(\sigma)q}{\varepsilon_0\omega}<\varepsilon_1\end{cases} \tag{S16}$$

and

$$\varphi_1=\arctan\left(\frac{\varepsilon_1-\frac{\Im(\sigma)q}{\varepsilon_0\omega}}{\sqrt{-\varepsilon_2^t\varepsilon_2^z}}\right),\quad \varphi_2=\arctan\left(\frac{\varepsilon_3}{\sqrt{-\varepsilon_2^t\varepsilon_2^z}}\right) \tag{S17}$$

Eq. S(16) can be further simplified to be:

$$q=\begin{cases}\frac{1}{d}\sqrt{-\frac{\varepsilon_2^z}{\varepsilon_2^t}}\left[m'\pi+\varphi_1+\varphi_2\right] & \varepsilon_2^t<0,\varepsilon_2^z>0\\ \frac{1}{d}\sqrt{-\frac{\varepsilon_2^z}{\varepsilon_2^t}}\left[m'\pi-\varphi_1-\varphi_2\right] & \varepsilon_2^t>0,\varepsilon_2^z<0\end{cases} \tag{S18}$$

Where $m'=-1, 0, 1, 2, \cdots$ this transcendental equation should be solved numerically.

2) SP[3] modes when $\varepsilon_2^t>0,\varepsilon_2^z>0$

When $\varepsilon_2^t>0,\varepsilon_2^z>0$, $k_2^z$ become imaginary and the HP[3] modes don't exist, but the SP[3] modes from graphene arise. To reduce the derivation, we use the dispersion result inherited from the reference (*4*) as follow:

$$e^{\kappa_2^z d}\left(\frac{\varepsilon_3}{\kappa_3^z}+\frac{\varepsilon_2^t}{\kappa_2^z}\right)\left(\frac{\varepsilon_1}{\kappa_1^z}+\frac{\varepsilon_2^t}{\kappa_2^z}+i\frac{\sigma}{\omega\varepsilon_0}\right)=e^{-\kappa_2^z d}\left(\frac{\varepsilon_3}{\kappa_3^z}-\frac{\varepsilon_2^t}{\kappa_2^z}\right)\left(\frac{\varepsilon_1}{\kappa_1^z}-\frac{\varepsilon_2^t}{\kappa_2^z}+i\frac{\sigma}{\omega\varepsilon_0}\right) \tag{S19}$$

$$\kappa_j^z=\sqrt{\varepsilon_j^t\left(\frac{q^2}{\varepsilon_j^z}-\frac{\omega^2}{c^2}\right)} \tag{S20}$$

After algebra, we can obtain:

$$e^{2\kappa_2^z d} = \frac{(\varepsilon_3 - \sqrt{\varepsilon_2^t \varepsilon_2^z})(\varepsilon_1 - \sqrt{\varepsilon_2^t \varepsilon_2^z} + i\frac{\sigma q}{\omega \varepsilon_0})}{(\varepsilon_3 + \sqrt{\varepsilon_2^t \varepsilon_2^z})(\varepsilon_1 + \sqrt{\varepsilon_2^t \varepsilon_2^z} + i\frac{\sigma q}{\omega \varepsilon_0})} \tag{S21}$$

This is also a transcendental equation and should be solved numerically. When the permittivity of the α-$MoO_3$ at a specific direction is large enough (an empirical value $\sqrt{\varepsilon_2^t \varepsilon_2^z} > 1$ for $d = 100$ nm), the evanescence wave will not penetrate the $SiO_2$ substrate, so we can ignore the effect of the $SiO_2$ substrate, and the dispersion relation of the $SP^3$ modes reduce to:

$$\frac{\varepsilon_1}{\kappa_1^z} + \frac{\varepsilon_2^t}{\kappa_2^z} + i\frac{\sigma}{\omega \varepsilon_0} = 0 \tag{S22}$$

$$\kappa_1^z \simeq q, \quad \kappa_2^z \simeq q\sqrt{\frac{\varepsilon_2^t}{\varepsilon_2^z}} \tag{S23}$$

After algebra and substituting the approximation of graphene conductivity Eq. (S6) in the mid-infrared region, we can obtain:

$$q = \frac{i\omega\varepsilon_0\left(\varepsilon_1 + \sqrt{\varepsilon_2^t \varepsilon_2^z}\right)}{\sigma} \simeq \frac{\pi\varepsilon_0 \hbar^2 \omega^2 \left(\varepsilon_1 + \sqrt{\varepsilon_2^t \varepsilon_2^z}\right)}{e^2 E_F} \tag{S24}$$

By combining Eq. (S18) and (S24), we obtain the dispersion relation of the graphene/α-$MoO_3$/$SiO_2$ heterostructure as shown in Eq.(1) of the main text. However, to show the excellent agreements between the analytical results with FDTD simulation results or pseudo-colored mapping of the imaginary parts of $r_p$, Eq. (S18) together with (S21) are used for the dispersion calculation along the $x$-axis ([100]) or $y$-axis ([001]) directions, and the IFCs calculations at the specific frequencies.

**S4. Wavelength tunability dependent on the Fermi level**

Here, we calculate the in-plane momentums along the $x$-axis ([100]) or $y$-axis ([001]) directions for frequencies of 623 and 925 $cm^{-1}$ by solving Eq. (S18) and (S21), then plot the corresponding wavelengths on the Fermi level as shown in Figure S3. Note that $m$ = 1 $HP^3$ mode for 623 $cm^{-1}$ along [100] direction is plotted because the wavelengths of the corresponding mode of $m$ = 0 are relatively large (~ 5 μm).

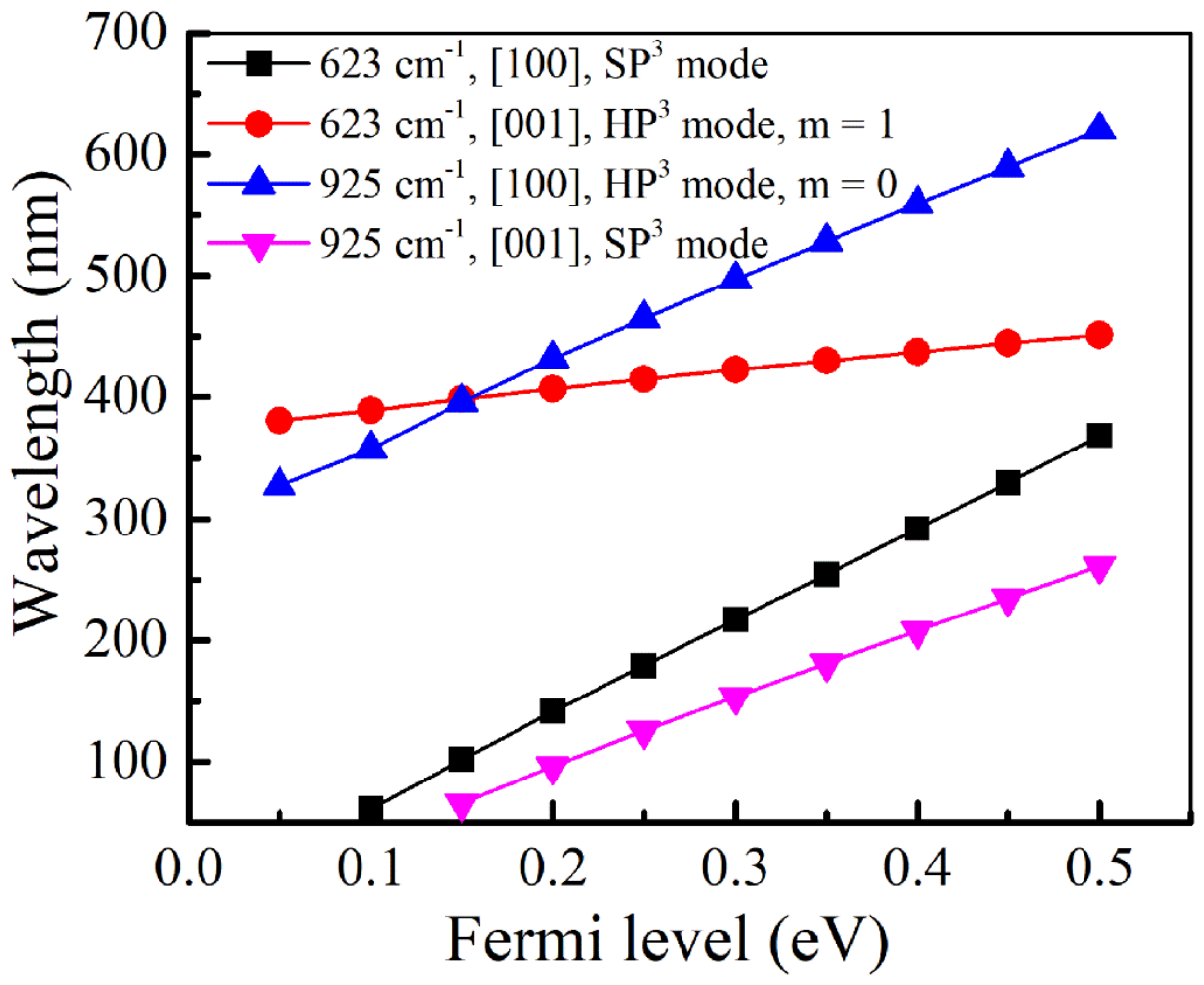

**Figure S3**. Wavelength tunability of the $HP^3$ and $SP^3$ modes of the graphene/α-$MoO_3$/$SiO_2$ structure, exhibiting a linear relationship with the Fermi level of the graphene.

**S5. Evolutions of the IFCs without graphene and with graphene at the varying Fermi level**

Here, Figure S4 shows the evolutions of the in-plane iso-frequency contours (IFCs) without graphene and with graphene at the Fermi levels of 0.15, 0.25, and 0.35 eV, at three typical frequencies of 623, 925, and 1000 $cm^{-1}$. The topological transition from open hyperbolic to bow-tie-like have been shown at the frequencies of 623 and 925 $cm^{-1}$, respectively. No topological transition occurs at the frequency of 1000 $cm^{-1}$.

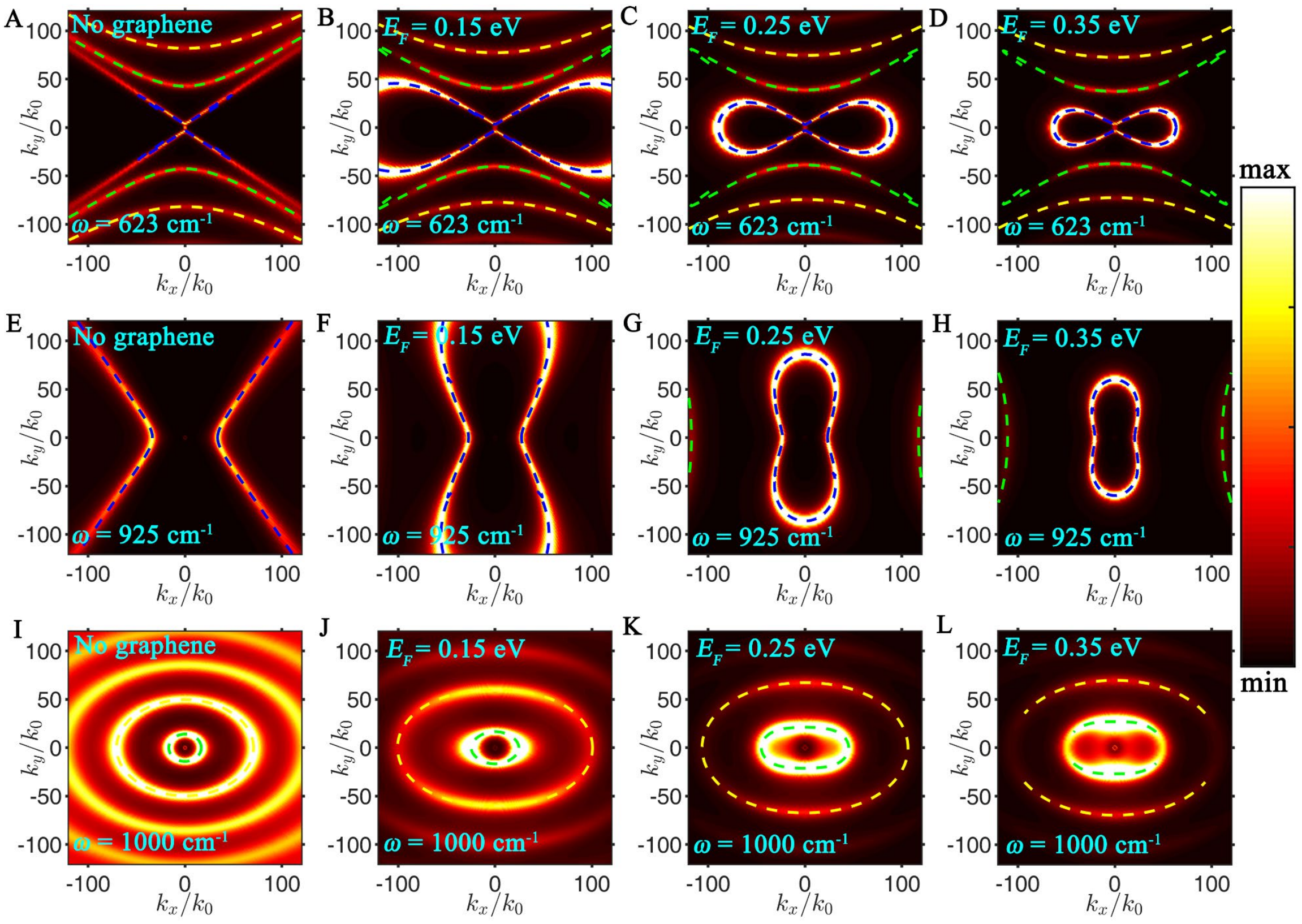

**Figure S4**. Evolutions of the IFCs in the graphene/α-$MoO_3$/$SiO_2$ structure without graphene and with graphene at the varying Fermi level of graphene. The pseudo-colored maps of the imaginary parts of $r_p$ in (A-D), (E-H), and (I-L) are calculated at the frequencies of 623, 925, and 1000 $cm^{-1}$, respectively. The blue dashed lines are the calculated $HP^3$ modes of $m = 0$ and $SP^3$ modes, and the green and yellow dashed lines are the $HP^3$ modes of $m$=1 and $m = 2$, respectively. The thickness of the α-$MoO_3$ slab is 100 nm.

## S6. Chiral excitation of waveguides with 3D hyperboloid or ellipsoid IFCs

Here, Figure S3 shows the highly asymmetric magnetic field spectral-spatial distribution of circular dipoles $\mathbf{p} = [1, 0, -i]$ and $\mathbf{p} = [0, 1, -i]$, calculated following the reference (5). The black solid lines in Figure S3(A, E), (B, F), and (C, G) are the three kinds of in-plane IFCs of 3D hyperboloid waveguide, namely hyperbolic with the open direction along $k_x$ or $k_y$ axis, and elliptical, respectively. The same elliptical in-plane IFCs in Figure S3 (D, H) are plotted for a 3D ellipsoid waveguide. The black arrows illustrate a predominant direction of the in-plane momentums of the corresponding dipoles, and the blue arrows illustrate the directions of the Poynting vectors **S** corresponding to the specific momentums of the waveguide modes. The components' directions ( $\pm x$ or $\pm y$ ) of momentum and Poynting vector are the same for the waveguide with 3D ellipsoid IFCs and be opposite for the 3D hyperboloid waveguide with elliptical in-plane IFCs, which means the chiral excitation directions of these two kinds of modes will be definitely opposite as shown in Figure S5 (C, D) and (G, H). Interestingly, the components' directions of momentum and Poynting vector can be the same or opposite for 3D hyperboloid waveguide with hyperbolic in-plane IFCs as shown in Figure S3 (A, E) and (B, F). Specifically, the $x$ components of momentum and the Poynting vector in Figure S3A are the same of both negative, but the $y$ component of the momentum is positive and opposite from the negative $y$ component of the Poynting vector.

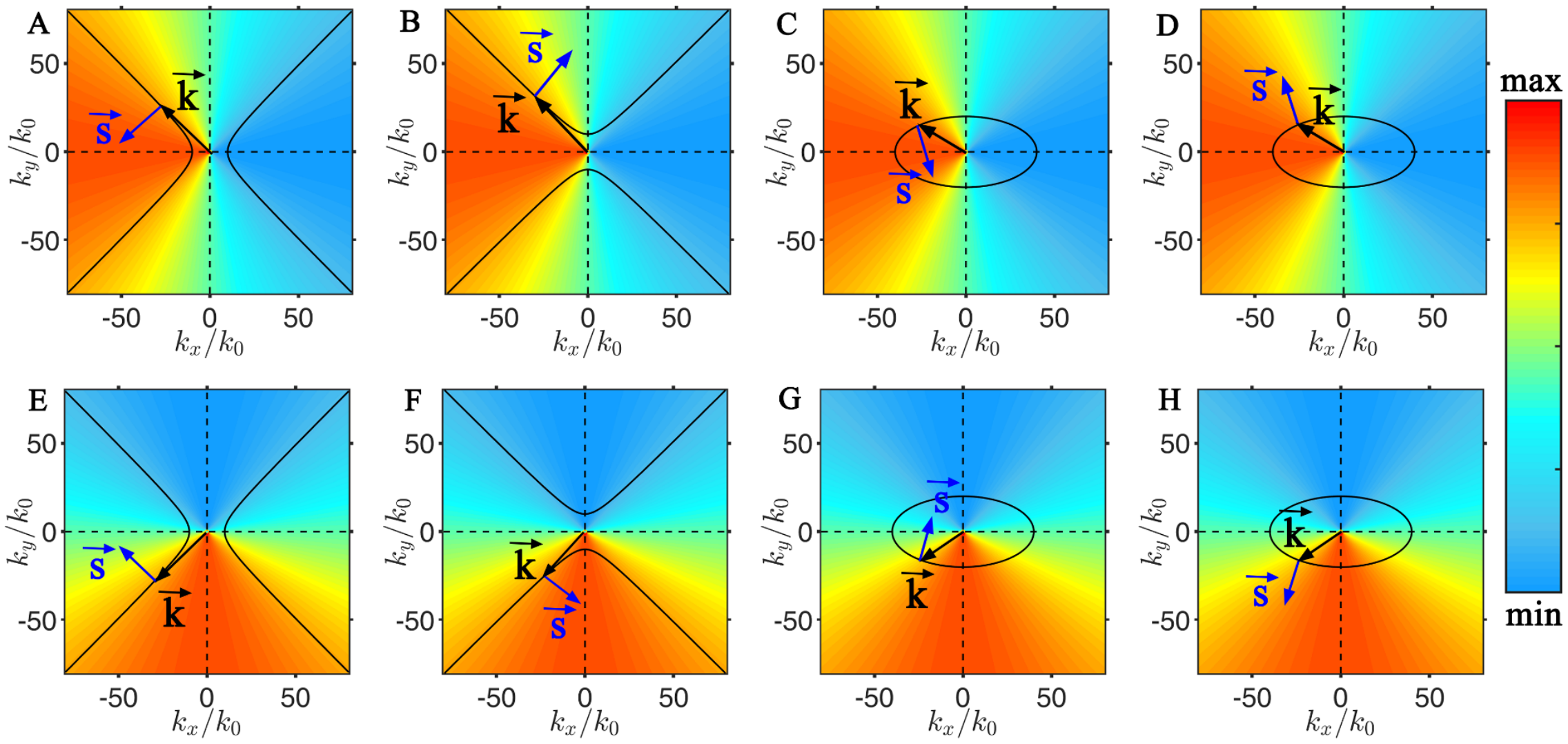

**Figure S5**. The magnetic field distribution of circular dipoles in ($k_x$, $k_y$) plane by spectral-spatial decomposition. (A-D) and (E-H) correspond to dipoles **p** = [1, 0, $-i$] and **p**=[0, 1, $-i$], respectively. The black solid lines in (A, E), (B, F), and (C, G) are three kinds of in-plane IFCs of waveguide with 3D hyperboloid IFCs. The black solid lines in (D, H) are elliptical in-plane IFCs of waveguide with 3D ellipsoid

IFCs. The black and blue arrows indicate the corresponding directions of momentums (**k**) and Poynting vectors (**S**).

## S7. Electrically tunable quantum spin Hall effects by excitation of circular dipole p = [0, 1, −*i*]

Here, Figure S4A and S4E show the Re($E_z$) and Poynting vector magnitude distribution of a 100 nm α-$MoO_3$ slab at $SiO_2$ substrate at the frequency of 623 $cm^{-1}$, excited by a circular dipole **p** = [0, 1, −*i*]. As shown in Figure S3(E-H), the magnetic field of the dipole is much higher in the region of $k_y < 0$. This asymmetric distribution leads to the predominant excitation of wave propagation along $y < 0$ for the frequency of 623 $cm^{-1}$. However, as shown in Figure S4I and S4M, the predominant excitation towards $y>0$ for the frequency of 925 $cm^{-1}$ is obvious, which should be attributed to the opposite directions of momentum and Poynting vector as shown in Fig S3E and Figure 2M of the main text. As the increase of Fermi levels of graphene, the predominant energy flow switches from $y > 0$ to $y < 0$ direction for frequency of 925 $cm^{-1}$, but remains towards $y < 0$ direction for frequency of 623 $cm^{-1}$. This exotic phenomenon should be attributed to the fact that the $SP^3$ modes gradually dominate the wave propagation at the interface as the increase of Fermi level.

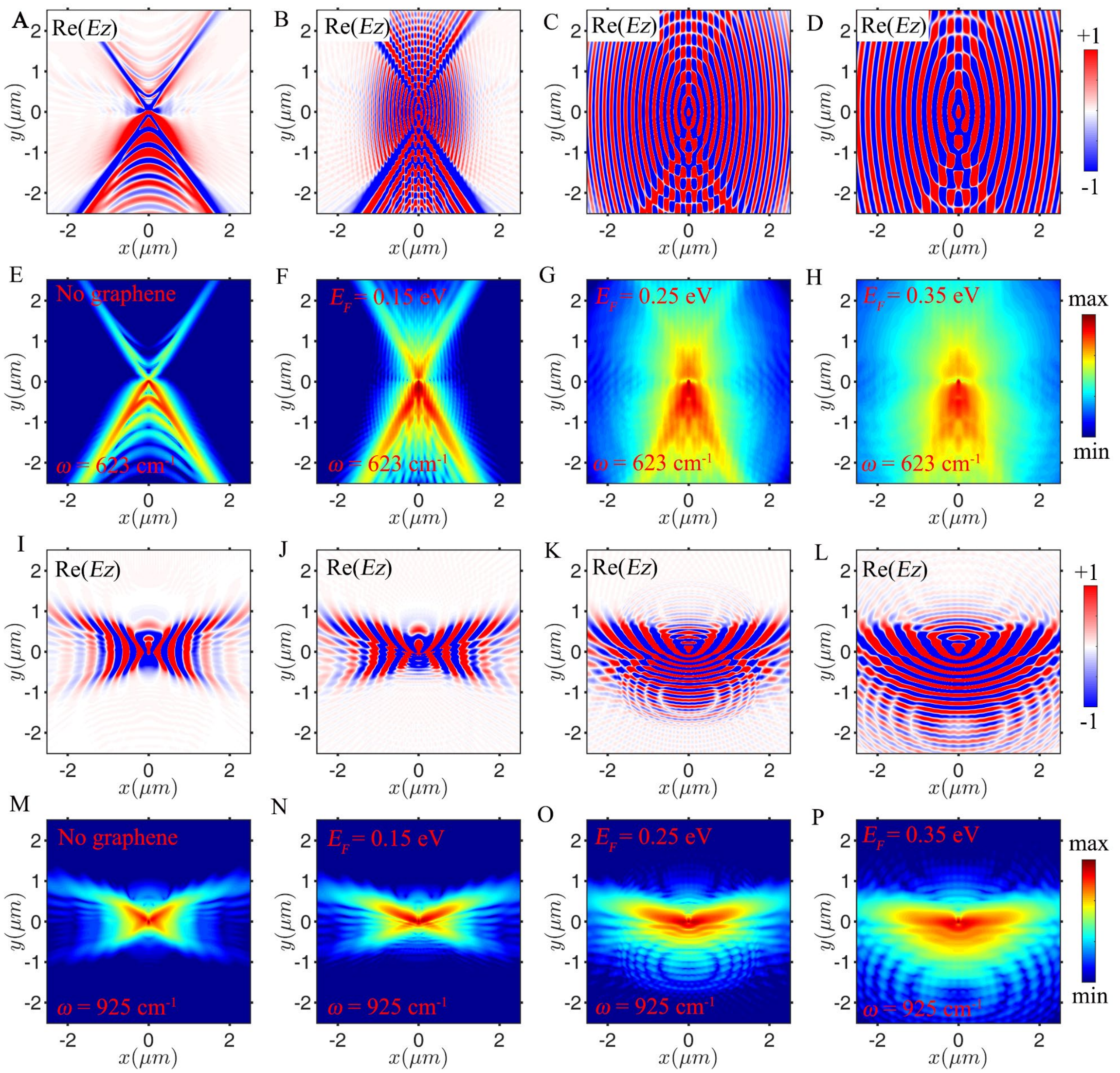

**Figure S6**. Electrically tunable spin Hall effects of plasmon-phonon polaritons in the graphene/α-$MoO_3$/$SiO_2$ structure. The excitation dipole **p** = [0, 1, −*i*] was located 60-nm-above the graphene. (A-D, I-L) Calculated real parts of *z*-components of the electric field distribution. (E-H, M-P) Corresponding Poynting vector magnitude without graphene and with graphene at the varying Fermi level. The frequency for (A-H) is 623 $cm^{-1}$, while for (I-P) is 925 $cm^{-1}$. The thickness of the α-$MoO_3$ slab is 100 nm.

## S8. Electrically tunable quantum spin Hall effect at the frequency of 1000 $cm^{-1}$

At frequency of 1000 $cm^{-1}$, the α-$MoO_3$ slab exhibit an elliptical in-plane IFCs, which lead to the opposite direction of momentum and Poynting vector for both dipole **p** = [1, 0, −*i*] and **p** = [0, 1, $-i$] excitation as shown in Figure S3(C, G). Figure S5(A, E) and S5(I, M) shows the field distribution of Re($E_z$) and Poynting vectors of the 100 nm α-$MoO_3$ slab at frequency of 1000 $cm^{-1}$ by excitation of circular dipole **p** = [1, 0, $-i$] and **p** = [0, 1, $-i$], respectively. As expected, the energy flow directions are both opposite to the momentum. However, as the increase of Fermi level of the graphene, the energy flow directions switch to be the same with the direction of momentum in the graphene/α-$MoO_3$ heterostructure, showing Fermi-levels-dependent quantum spin Hall effects. This should be attributed to the $SP^3$-dominated wave propagation at the interface when a large Fermi level is applied.

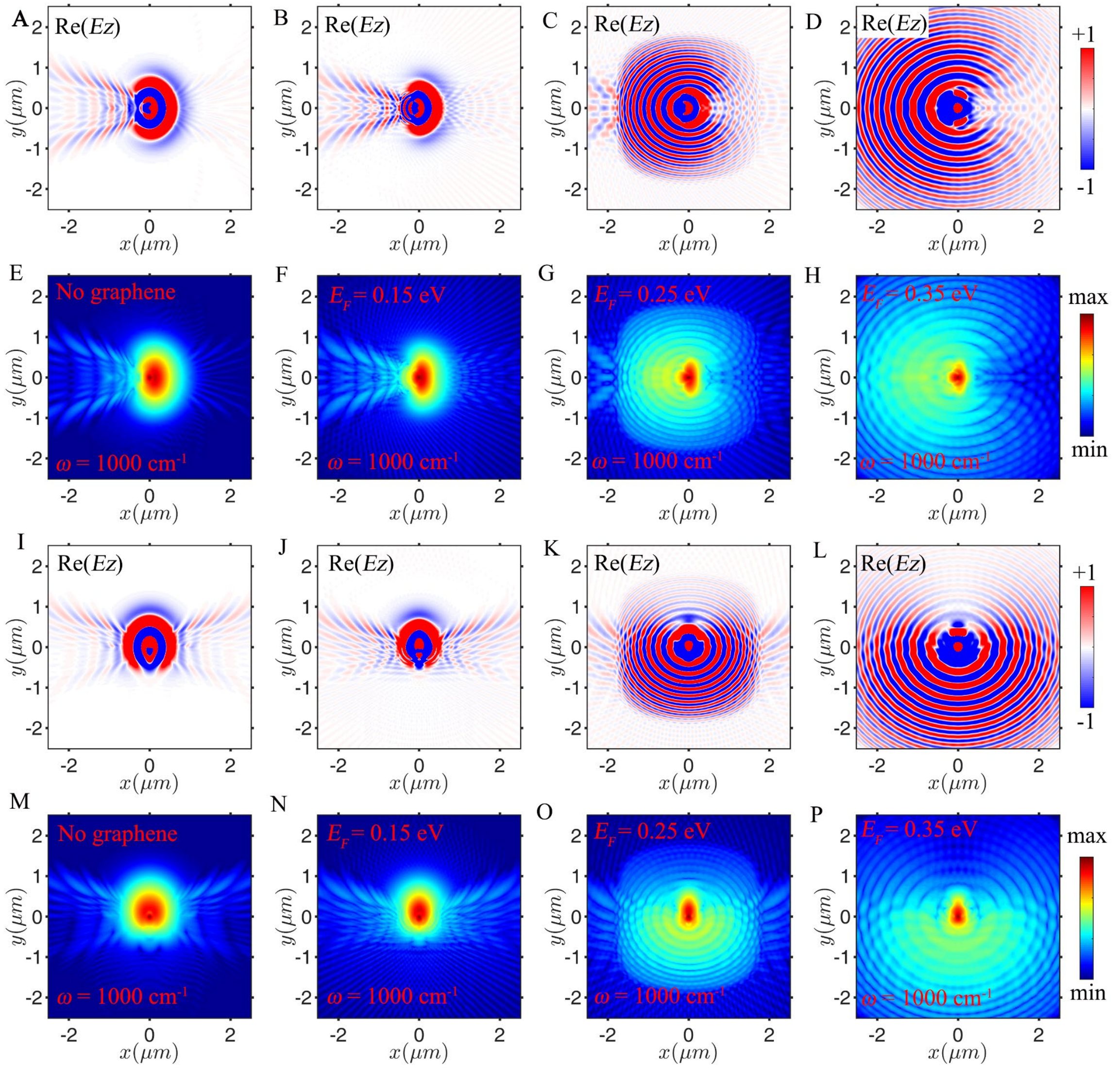

**Figure S7**. Electrically tunable quantum spin Hall effect of plasmon-phonon polaritons in the graphene/α-$MoO_3/SiO_2$ structure. For (A-H), the circular dipole $\mathbf{p} = [1, 0, -i]$ was used for excitation, while the dipole $\mathbf{p} = [0, 1, -i]$ was used for (I-P). The dipoles were located 60-nm-above of the graphene. (A-D, I-L) Calculated real parts of $z$-components of the electric field distribution. (E-H, M-P) Log scale plots of the corresponding Poynting vector magnitudes without graphene and with graphene at the varying Fermi level. The calculations are performed at the frequency of 1000 $cm^{-1}$. The thickness of the α-$MoO_3$ slab is 100 nm.